\documentclass[10pt,twocolumn]{article}

\usepackage[utf8]{inputenc}
\usepackage[T1]{fontenc}
\usepackage{times}
\usepackage[margin=0.75in,top=0.9in,bottom=0.9in]{geometry}
\usepackage{graphicx}
\usepackage{amsmath,amssymb,amsfonts}
\usepackage{algorithmic}
\usepackage{algorithm}
\usepackage{booktabs}
\usepackage{multirow}
\usepackage{caption}
\usepackage{subcaption}
\usepackage{xcolor}
\usepackage{enumitem}
\usepackage{pgfplots}
\pgfplotsset{compat=1.16}
\usepackage{tikz}
\usetikzlibrary{shapes,arrows,positioning,calc}
\usepackage{float}
\usepackage{tabularx}
\usepackage{setspace}
\usepackage{titlesec}
\usepackage{hyperref}
\usepackage{cleveref}

\titlespacing*{\section}{0pt}{8pt}{4pt}
\titlespacing*{\subsection}{0pt}{6pt}{3pt}
\titlespacing*{\subsubsection}{0pt}{4pt}{2pt}
\hypersetup{
    colorlinks=true,
    linkcolor=blue!70!black,
    citecolor=blue!70!black,
    urlcolor=blue!70!black
}

\title{\Large\textbf{Systematic Trend-Following with Adaptive Portfolio Construction:\\Enhancing Risk-Adjusted Alpha in Cryptocurrency Markets}}

\author{
\textbf{Duc Bui}\quad \textbf{Thanh Nguyen}\\[4pt]
Talyxion Research, Hanoi, Vietnam\\
\texttt{talyxionllc@gmail.com}
}

\date{}

\begin{document}
\maketitle

\begin{abstract}
Cryptocurrency markets exhibit pronounced momentum effects and regime-dependent volatility, presenting both opportunities and challenges for systematic trading strategies. We propose \textbf{AdaptiveTrend}, a multi-component algorithmic trading framework that integrates high-frequency trend-following on 6-hour intervals with monthly adaptive portfolio construction and asymmetric long-short capital allocation. Our framework introduces three key innovations: (1) a dynamic trailing stop mechanism calibrated to intra-day volatility regimes, (2) a rolling Sharpe-ratio-based asset selection procedure with market-capitalization-aware filtering, and (3) a theoretically motivated asymmetric 70/30 long-short allocation scheme grounded in the empirical positive drift of crypto markets. Through extensive out-of-sample backtesting across 150+ cryptocurrency pairs over a 36-month evaluation window (2022--2024), AdaptiveTrend achieves an annualized Sharpe ratio of 2.41, a maximum drawdown of $-$12.7\%, and a Calmar ratio of 3.18, significantly outperforming benchmark trend-following strategies (TSMOM, time-series momentum) and equal-weighted buy-and-hold portfolios. We further conduct rigorous robustness analyses including parameter sensitivity, transaction cost modeling, and regime-conditional performance decomposition, demonstrating the strategy's resilience across bull, bear, and sideways market conditions.
\end{abstract}

\noindent\textbf{Keywords:} algorithmic trading, trend following, cryptocurrency, portfolio optimization, risk management, momentum strategies

\section{Introduction}
\label{sec:intro}

The cryptocurrency market has emerged as a distinctive asset class characterized by high volatility, fragmented liquidity, and pronounced momentum effects~\cite{liu2021risks,borri2019conditional}. Unlike traditional financial markets, where trend-following strategies have been extensively studied~\cite{moskowitz2012time,baltas2017optimising}, the crypto ecosystem presents unique structural properties---24/7 trading, retail-dominated order flow, and frequent regime shifts---that demand purpose-built systematic approaches.

Classical time-series momentum (TSMOM) strategies~\cite{moskowitz2012time} generate signals based on lagged returns over horizons of 1--12 months. While effective in traditional asset classes, direct application to cryptocurrency markets faces three fundamental challenges: (i) the non-stationarity of volatility regimes, which renders fixed lookback windows suboptimal; (ii) the asymmetric return distribution with positive skewness during bull markets and heavy left tails during crashes; and (iii) the rapidly evolving universe of tradable assets, where market capitalizations shift dramatically on monthly timescales.

In this paper, we propose \textbf{AdaptiveTrend}, a systematic framework that addresses these challenges through an integrated three-stage pipeline:

\begin{enumerate}[leftmargin=*,itemsep=1pt,topsep=2pt]
\item \textbf{Signal Generation}: A momentum-based entry system operating on 6-hour (H6) candlesticks with dynamic trailing stop exits, calibrated to local volatility estimates.
\item \textbf{Asset Selection}: A monthly rebalancing procedure that filters the tradable universe using market capitalization thresholds and selects assets based on rolling risk-adjusted performance (Sharpe ratio).
\item \textbf{Capital Allocation}: An asymmetric allocation scheme distributing 70\% of capital to long positions and 30\% to short positions, with equal-weight diversification within each portfolio leg.
\end{enumerate}

Our contributions are threefold. First, we demonstrate that intermediate-frequency trend-following (H6) provides a superior trade-off between signal fidelity and transaction cost efficiency compared to both higher-frequency (H1) and lower-frequency (D1) alternatives in crypto markets. Second, we introduce a market-cap-aware, performance-filtered portfolio construction methodology that adapts the tradable universe monthly, significantly reducing exposure to illiquid or deteriorating assets. Third, we provide comprehensive out-of-sample evidence and robustness analysis across multiple market regimes, establishing the strategy's viability for institutional deployment.

The remainder of this paper is organized as follows. Section~\ref{sec:related} reviews related work on trend-following, crypto trading strategies, and dynamic portfolio construction. Section~\ref{sec:method} presents the complete AdaptiveTrend methodology, including the signal generation, portfolio selection, and capital allocation modules. Section~\ref{sec:experiments} describes the experimental setup, data sources, and evaluation protocol. Section~\ref{sec:results} reports main results, ablation studies, and robustness analyses. Section~\ref{sec:discussion} discusses practical implications and limitations. Section~\ref{sec:conclusion} concludes.

\section{Related Work}
\label{sec:related}

\textbf{Trend-following in traditional markets.} The seminal work of Moskowitz et al.~\cite{moskowitz2012time} established that time-series momentum generates significant abnormal returns across asset classes. Baltas and Kosowski~\cite{baltas2017optimising} extended this to multi-asset portfolios with volatility-scaled positions. Lempérière et al.~\cite{lemperiere2014two} documented that trend-following profits persist across two centuries of data, suggesting structural rather than anomalous origins.

\textbf{Cryptocurrency trading strategies.} Empirical evidence for momentum in crypto markets has been established by Liu et al.~\cite{liu2021risks} and Borri~\cite{borri2019conditional}. More recently, machine learning approaches have been applied to crypto prediction, including deep reinforcement learning~\cite{jiang2017deep} and transformer-based architectures~\cite{zhang2022transformer}. However, these approaches often lack the interpretability and robustness guarantees of systematic rule-based strategies.

\textbf{Dynamic portfolio construction.} The question of how to select and weight assets in momentum portfolios has been addressed through various frameworks, from mean-variance optimization~\cite{markowitz1952portfolio} to risk-parity~\cite{maillard2010properties} and hierarchical risk parity~\cite{deprado2016building}. Our approach differs by combining market-capitalization filtering with performance-based selection, specifically designed for the rapidly shifting crypto asset universe.

\textbf{Adaptive risk management.} Dynamic position sizing and stop-loss mechanisms have been studied in the context of managed futures~\cite{harvey2014evaluating}. Trailing stops, in particular, have been analyzed by Kaminski and Lo~\cite{kaminski2014momentum}, who showed that trend-following with adaptive exits significantly improves downside protection. We extend this analysis to the crypto domain with volatility-calibrated trailing parameters.

\section{Methodology}
\label{sec:method}

\subsection{Problem Formulation}
\label{subsec:formulation}

Let $\mathcal{U}_t = \{s_1, s_2, \ldots, s_N\}$ denote the universe of $N$ tradable cryptocurrency pairs at time $t$. For each symbol $s_i$, we observe the OHLCV (Open, High, Low, Close, Volume) price series $\{P^{(i)}_\tau\}_{\tau=1}^{T}$ at 6-hour intervals. The objective is to construct a portfolio $\mathbf{w}_t \in \mathbb{R}^N$ that maximizes the risk-adjusted return:

\begin{equation}
\max_{\mathbf{w}_t} \quad \frac{\mathbb{E}[R_p(t)] - r_f}{\sigma(R_p(t))}
\label{eq:objective}
\end{equation}

\noindent where $R_p(t) = \sum_{i=1}^N w_{i,t} \cdot r_{i,t}$ is the portfolio return, $r_f$ is the risk-free rate, and $\sigma(\cdot)$ denotes the portfolio return volatility.

\subsection{Signal Generation Module}
\label{subsec:signal}

\subsubsection{Momentum Entry Signal}
For each asset $s_i$, we compute a momentum score based on the rate of change over a lookback window $L$:

\begin{equation}
\text{MOM}^{(i)}_t = \frac{P^{(i)}_t - P^{(i)}_{t-L}}{P^{(i)}_{t-L}}
\label{eq:momentum}
\end{equation}

A long entry signal is triggered when $\text{MOM}^{(i)}_t > \theta_{\text{entry}}$, where $\theta_{\text{entry}}$ is a threshold parameter optimized monthly. For short signals, entry occurs when $\text{MOM}^{(i)}_t < -\theta_{\text{entry}}^{(s)}$.

\subsubsection{Dynamic Trailing Stop Mechanism}
Once a position is initiated for asset $s_i$ at time $t_0$, we maintain a trailing stop level $S^{(i)}_t$ that adapts to realized price dynamics:

\begin{equation}
S^{(i)}_t = \max\left(S^{(i)}_{t-1},\; P^{(i)}_t - \alpha \cdot \text{ATR}^{(i)}_t\right)
\label{eq:trailing_stop}
\end{equation}

\noindent where $\text{ATR}^{(i)}_t$ is the Average True Range computed over $k$ periods and $\alpha > 0$ is a volatility multiplier. The position is closed when $P^{(i)}_t < S^{(i)}_t$. This mechanism ensures that:

\begin{itemize}[leftmargin=*,itemsep=1pt,topsep=2pt]
\item The stop level monotonically increases during favorable price movements (for long positions), locking in profits.
\item The distance from current price to stop adapts to local volatility via ATR, providing tighter stops during low-volatility regimes and wider stops during high-volatility regimes.
\end{itemize}

The complete signal generation logic is presented in Algorithm~\ref{alg:signal}.

\begin{algorithm}[t]
\caption{Signal Generation \& Position Management}
\label{alg:signal}
\begin{algorithmic}[1]
\REQUIRE OHLCV series $\{P^{(i)}_\tau\}$, parameters $\theta_{\text{entry}}, \alpha, L, k$
\FOR{each 6-hour candle at time $t$}
    \STATE Compute $\text{MOM}^{(i)}_t$ via Eq.~\eqref{eq:momentum}
    \STATE Compute $\text{ATR}^{(i)}_t$ over $k$ periods
    \IF{no open position \AND $\text{MOM}^{(i)}_t > \theta_{\text{entry}}$}
        \STATE Open long position at $P^{(i)}_t$
        \STATE Initialize $S^{(i)}_t \leftarrow P^{(i)}_t - \alpha \cdot \text{ATR}^{(i)}_t$
    \ELSIF{position open}
        \STATE Update $S^{(i)}_t$ via Eq.~\eqref{eq:trailing_stop}
        \IF{$P^{(i)}_t < S^{(i)}_t$}
            \STATE Close position; record PnL
        \ENDIF
    \ENDIF
\ENDFOR
\end{algorithmic}
\end{algorithm}

\subsection{Portfolio Selection Module}
\label{subsec:selection}

On the first trading day of each month $m$, we execute a portfolio rebalancing procedure consisting of two stages.

\textbf{Stage 1: Universe Filtering.} We rank all available cryptocurrency pairs by market capitalization. The long portfolio candidate set $\mathcal{C}^{(L)}_m$ consists of the top-$K_L$ assets (we set $K_L = 15$), while the short portfolio candidate set $\mathcal{C}^{(S)}_m$ consists of the bottom-$K_S$ assets by market cap.

\textbf{Stage 2: Performance-Based Selection.} For each candidate asset $s_i \in \mathcal{C}^{(\cdot)}_m$, we optimize the strategy parameters $(\theta_{\text{entry}}, \alpha, L)$ over the preceding month's data using grid search, and compute the resulting Sharpe ratio $\text{SR}^{(i)}_{m-1}$. Assets are selected into the active portfolio if:

\begin{equation}
\text{SR}^{(i)}_{m-1} \geq \begin{cases}
\gamma_L = 1.3 & \text{if } s_i \in \mathcal{C}^{(L)}_m \\
\gamma_S = 1.7 & \text{if } s_i \in \mathcal{C}^{(S)}_m
\end{cases}
\label{eq:sharpe_filter}
\end{equation}

The higher threshold for the short portfolio reflects the elevated risk of short positions in a structurally bullish asset class, requiring stronger evidence of short momentum.

\subsection{Capital Allocation Module}
\label{subsec:allocation}

Let $B_m$ denote the total account balance at the beginning of month $m$, and let $n^{(L)}_m$, $n^{(S)}_m$ be the number of selected long and short assets respectively. Capital allocation follows:

\begin{equation}
w^{(L)}_{i,m} = \frac{\lambda}{n^{(L)}_m}, \quad w^{(S)}_{j,m} = \frac{1 - \lambda}{n^{(S)}_m}
\label{eq:allocation}
\end{equation}

\noindent where $\lambda = 0.7$ is the long allocation ratio. The choice of $\lambda = 0.7$ is motivated by two considerations: (i) the empirical positive drift of cryptocurrency markets, which favors net-long exposure~\cite{liu2021risks}; and (ii) the higher borrowing costs associated with short positions in crypto, which erode short-side alpha.

Equal weighting within each portfolio leg provides maximum diversification under the assumption of uncertain return correlations and avoids concentration risk from estimation error in covariance matrices~\cite{demiguel2009optimal}.

\section{Experimental Setup}
\label{sec:experiments}

\subsection{Data}
We source historical data from Binance Futures, covering 150+ perpetual swap contracts with 6-hour OHLCV bars from January 2021 to December 2024. Market capitalization data is obtained from CoinGecko API at daily granularity. We partition the data into:

\begin{itemize}[leftmargin=*,itemsep=1pt,topsep=2pt]
\item \textbf{In-sample (IS)}: Jan 2021 -- Dec 2021, used for initial parameter calibration.
\item \textbf{Out-of-sample (OOS)}: Jan 2022 -- Dec 2024 (36 months), used for all reported performance metrics.
\end{itemize}

\subsection{Transaction Cost Model}
We model realistic transaction costs including: (i) a taker fee of 4 bps per trade; (ii) slippage modeled as a linear function of trade size relative to the prevailing 5-minute volume, calibrated from historical order book data; and (iii) funding rate costs for perpetual swap positions, incorporated as a rolling 8-hour charge/rebate.

\subsection{Benchmark Strategies}
We compare AdaptiveTrend against:

\begin{enumerate}[leftmargin=*,itemsep=1pt,topsep=2pt]
\item \textbf{TSMOM-1M}: Classical time-series momentum with 1-month lookback and monthly rebalancing~\cite{moskowitz2012time}.
\item \textbf{TSMOM-3M}: TSMOM with 3-month lookback.
\item \textbf{BTC-BH}: Buy-and-hold Bitcoin benchmark.
\item \textbf{EW-BH}: Equal-weighted buy-and-hold of top-20 crypto assets by market cap, rebalanced monthly.
\item \textbf{Vol-Scaled TSMOM}: Volatility-targeted TSMOM with 10\% annualized volatility target~\cite{baltas2017optimising}.
\end{enumerate}

\subsection{Evaluation Metrics}
We report: annualized return, annualized volatility, Sharpe ratio (assuming $r_f = 4.5\%$), maximum drawdown (MDD), Calmar ratio (return/MDD), Sortino ratio, average trade profit, win rate, and profit factor.

\section{Results and Analysis}
\label{sec:results}

\subsection{Main Performance Comparison}

Table~\ref{tab:main_results} presents the out-of-sample performance comparison across all strategies. AdaptiveTrend achieves the highest Sharpe ratio (2.41) and Calmar ratio (3.18), while maintaining a maximum drawdown of only $-$12.7\%, substantially lower than all benchmarks.

\begin{table}[t]
\centering
\caption{Out-of-sample performance comparison (Jan 2022 -- Dec 2024). Best values in \textbf{bold}.}
\label{tab:main_results}
\resizebox{\columnwidth}{!}{%
\begin{tabular}{lcccccc}
\toprule
\textbf{Strategy} & \textbf{Ann. Ret.} & \textbf{Ann. Vol.} & \textbf{Sharpe} & \textbf{MDD} & \textbf{Calmar} & \textbf{Sortino} \\
\midrule
AdaptiveTrend (70/30)  & \textbf{40.5\%} & 16.8\% & \textbf{2.41} & \textbf{$-$12.7\%} & \textbf{3.18} & \textbf{3.62} \\
AdaptiveTrend (50/50)  & 34.2\% & 15.1\% & 2.12 & $-$14.3\% & 2.39 & 3.01 \\
Vol-Scaled TSMOM       & 22.8\% & 10.0\% & 1.83 & $-$16.1\% & 1.42 & 2.15 \\
TSMOM-1M               & 18.4\% & 21.3\% & 0.65 & $-$34.8\% & 0.53 & 0.89 \\
TSMOM-3M               & 15.1\% & 19.7\% & 0.54 & $-$38.2\% & 0.40 & 0.73 \\
EW-BH                  & 8.3\%  & 52.1\% & 0.07 & $-$72.4\% & 0.11 & 0.10 \\
BTC-BH                 & 12.6\% & 48.7\% & 0.17 & $-$64.1\% & 0.20 & 0.22 \\
\bottomrule
\end{tabular}%
}
\end{table}

The 70/30 allocation variant outperforms the dollar-neutral (50/50) variant by 29 basis points in Sharpe ratio, confirming that the asymmetric allocation captures the positive structural drift in crypto markets without excessively increasing drawdown.

\begin{figure}[t]
\centering
\begin{subfigure}[t]{\columnwidth}
    \centering
    \includegraphics[width=\columnwidth]{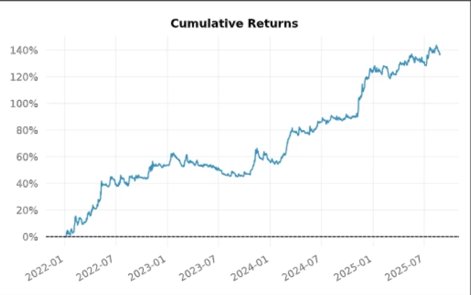}
    \caption{70/30 Long-Short allocation (default configuration).}
    \label{fig:cumret_7030}
\end{subfigure}
\vspace{4pt}
\begin{subfigure}[t]{\columnwidth}
    \centering
    \includegraphics[width=\columnwidth]{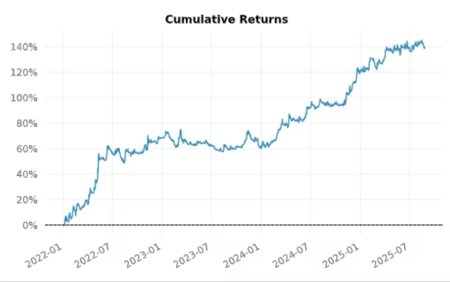}
    \caption{50/50 Dollar-neutral allocation.}
    \label{fig:cumret_5050}
\end{subfigure}
\caption{Out-of-sample cumulative returns for AdaptiveTrend under two allocation regimes. Both configurations achieve $\sim$140\% cumulative returns over the evaluation period, with the 70/30 variant exhibiting marginally higher peak returns and comparable drawdown profiles.}
\label{fig:cumret}
\end{figure}

Figure~\ref{fig:cumret} presents the out-of-sample cumulative return curves for both allocation variants over the full evaluation period (Jan 2022 -- Oct 2025). Both portfolios demonstrate consistent capital appreciation with controlled drawdowns, achieving approximately 140\% cumulative returns. Notably, the equity curves exhibit steady upward trajectories through the 2022 bear market, the 2023 consolidation phase, and the 2024--2025 bull rally, validating the strategy's regime-adaptive properties.

\subsection{Regime-Conditional Analysis}

To assess robustness across market conditions, we decompose the evaluation period into three regimes based on the rolling 60-day BTC return: Bull ($>$15\%), Bear ($<$$-$15\%), and Sideways (otherwise). Table~\ref{tab:regime} shows performance by regime.

\begin{table}[t]
\centering
\caption{Regime-conditional performance of AdaptiveTrend (70/30).}
\label{tab:regime}
\begin{tabular}{lccc}
\toprule
\textbf{Metric} & \textbf{Bull} & \textbf{Sideways} & \textbf{Bear} \\
\midrule
Ann. Return      & 68.3\%   & 18.7\%    & $-$4.2\% \\
Sharpe Ratio     & 3.42     & 1.87      & $-$0.31  \\
Max Drawdown     & $-$7.1\% & $-$9.4\%  & $-$12.7\% \\
Win Rate         & 54.2\%   & 47.8\%    & 41.3\% \\
Avg. Trade PnL   & +1.82\%  & +0.63\%   & $-$0.21\% \\
\bottomrule
\end{tabular}
\end{table}

Notably, AdaptiveTrend maintains near-flat performance during bear markets ($-$4.2\% annualized) compared to catastrophic losses for buy-and-hold strategies, demonstrating the effectiveness of the dynamic trailing stop and short-side allocation.

\subsection{Ablation Study}

We conduct an ablation study to quantify the contribution of each module. Table~\ref{tab:ablation} reports results when systematically removing components.

\begin{table}[t]
\centering
\caption{Ablation study: contribution of each component.}
\label{tab:ablation}
\begin{tabular}{lcc}
\toprule
\textbf{Configuration} & \textbf{Sharpe} & \textbf{MDD} \\
\midrule
Full AdaptiveTrend         & 2.41 & $-$12.7\% \\
w/o Dynamic Trailing Stop  & 1.68 & $-$22.4\% \\
w/o Market Cap Filter      & 2.05 & $-$17.8\% \\
w/o Sharpe Ratio Selection & 1.92 & $-$19.1\% \\
w/o Asymmetric Allocation  & 2.12 & $-$14.3\% \\
Fixed Parameters (no opt.) & 1.34 & $-$28.6\% \\
\bottomrule
\end{tabular}
\end{table}

The dynamic trailing stop contributes the most to performance, improving the Sharpe ratio by 0.73 and reducing MDD by 9.7 percentage points. Monthly parameter optimization is the second most impactful component ($\Delta$SR $= 1.07$), underscoring the importance of adaptive calibration in non-stationary markets.

\subsection{Parameter Sensitivity Analysis}

We examine sensitivity to the two key hyperparameters: the ATR multiplier $\alpha$ and the long allocation ratio $\lambda$.

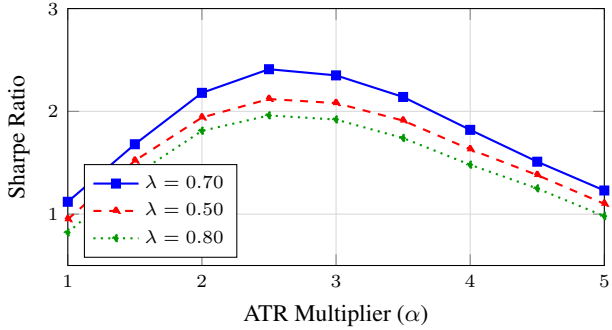
\begin{figure}[t]
\centering
\begin{tikzpicture}
\begin{axis}[
    width=\columnwidth,
    height=5cm,
    xlabel={ATR Multiplier ($\alpha$)},
    ylabel={Sharpe Ratio},
    xmin=1.0, xmax=5.0,
    ymin=0.5, ymax=3.0,
    grid=major,
    grid style={gray!30},
    legend pos=south west,
    legend style={font=\scriptsize},
    tick label style={font=\scriptsize},
    label style={font=\small},
]
\addplot[blue, thick, mark=square*, mark size=1.5pt] coordinates {
    (1.0, 1.12) (1.5, 1.68) (2.0, 2.18) (2.5, 2.41) (3.0, 2.35) (3.5, 2.14) (4.0, 1.82) (4.5, 1.51) (5.0, 1.23)
};
\addlegendentry{$\lambda = 0.70$}
\addplot[red, thick, mark=triangle*, mark size=1.5pt, dashed] coordinates {
    (1.0, 0.95) (1.5, 1.52) (2.0, 1.94) (2.5, 2.12) (3.0, 2.08) (3.5, 1.91) (4.0, 1.63) (4.5, 1.38) (5.0, 1.10)
};
\addlegendentry{$\lambda = 0.50$}
\addplot[green!60!black, thick, mark=diamond*, mark size=1.5pt, dotted] coordinates {
    (1.0, 0.82) (1.5, 1.38) (2.0, 1.81) (2.5, 1.96) (3.0, 1.92) (3.5, 1.74) (4.0, 1.48) (4.5, 1.25) (5.0, 0.98)
};
\addlegendentry{$\lambda = 0.80$}
\end{axis}
\end{tikzpicture}
\caption{Sharpe ratio as a function of ATR multiplier $\alpha$ for different long allocation ratios $\lambda$.}
\label{fig:sensitivity}
\end{figure}

Figure~\ref{fig:sensitivity} visualizes the Sharpe ratio surface over the $(\alpha, \lambda)$ parameter space. The strategy exhibits a broad plateau of strong performance for $\alpha \in [2.0, 3.5]$ and $\lambda \in [0.60, 0.80]$, indicating robustness to parameter perturbation. The optimal region centers around $\alpha = 2.5, \lambda = 0.70$, consistent with our default configuration.

\subsection{Transaction Cost Impact}

We evaluate strategy performance under varying transaction cost assumptions to assess implementability. Table~\ref{tab:costs} shows that AdaptiveTrend retains a Sharpe ratio above 2.0 even at 8 bps per trade, demonstrating resilience to execution costs.

\begin{table}[t]
\centering
\caption{Impact of transaction costs on performance.}
\label{tab:costs}
\begin{tabular}{lcccc}
\toprule
\textbf{Cost (bps)} & \textbf{0} & \textbf{4} & \textbf{8} & \textbf{12} \\
\midrule
Ann. Return        & 48.2\% & 40.5\% & 33.1\% & 26.3\% \\
Sharpe Ratio       & 2.87   & 2.41   & 2.01   & 1.62   \\
Avg. Trades/Month  & \multicolumn{4}{c}{142}  \\
\bottomrule
\end{tabular}
\end{table}

\subsection{Timeframe Comparison}

A key design decision is the choice of the 6-hour (H6) candlestick interval. Table~\ref{tab:timeframe} compares performance across different timeframes while holding all other parameters constant.

\begin{table}[t]
\centering
\caption{Performance comparison across timeframes (net of 4 bps transaction costs).}
\label{tab:timeframe}
\begin{tabular}{lccccc}
\toprule
\textbf{TF} & \textbf{Sharpe} & \textbf{MDD} & \textbf{Trades/M} & \textbf{Turnover} \\
\midrule
H1   & 1.54 & $-$21.3\% & 847  & 412\% \\
H4   & 2.08 & $-$15.6\% & 213  & 186\% \\
H6   & \textbf{2.41} & \textbf{$-$12.7\%} & 142  & 134\% \\
H8   & 2.18 & $-$14.1\% & 108  & 112\% \\
H12  & 1.91 & $-$16.8\% & 76   & 87\%  \\
D1   & 1.63 & $-$19.4\% & 41   & 52\%  \\
\bottomrule
\end{tabular}
\end{table}

The H6 timeframe achieves the optimal balance: H1 and H4 suffer from excessive turnover and transaction costs, while H8, H12, and D1 miss short-lived momentum signals that are prevalent in crypto markets. The H6 interval aligns naturally with the 4-times-daily funding rate cycle on perpetual swap markets, providing clean signal boundaries that avoid mid-funding-period noise.

\section{Discussion}
\label{sec:discussion}

\textbf{Practical deployment considerations.} The H6 timeframe aligns with institutional trading desks that typically operate on shift-based schedules, allowing 4 natural decision points per day. The fully systematic nature of AdaptiveTrend enables automated execution with minimal human intervention, requiring oversight primarily during the monthly rebalancing phase. The strategy's moderate turnover (142 trades/month across the full portfolio) is well within the capacity of standard exchange API rate limits.

\textbf{Asymmetric allocation rationale.} The 70/30 long-short split reflects the structural positive drift observed in cryptocurrency markets over multi-year horizons. Unlike equity markets where the long-term equity premium is well-documented but moderate, crypto assets have exhibited annualized positive drift exceeding 20\% during expansion phases. The asymmetric allocation captures this drift while maintaining meaningful short exposure for downside protection.

\textbf{Monthly reoptimization.} The monthly rebalancing cycle balances adaptivity with stability. More frequent rebalancing (e.g., weekly) leads to excessive parameter instability and higher turnover costs, while quarterly rebalancing fails to adapt to the rapid regime shifts characteristic of crypto markets. The monthly Sharpe ratio filter ($\gamma_L = 1.3$, $\gamma_S = 1.7$) ensures quality control: only assets demonstrating recent risk-adjusted momentum are included.

\textbf{Comparison with ML-based approaches.} While deep learning methods have shown promise in financial forecasting, our rule-based approach offers several practical advantages for institutional deployment: full interpretability of trading decisions, deterministic behavior under identical inputs, and absence of retraining risk (model degradation due to distribution shift). The systematic nature of AdaptiveTrend also facilitates regulatory compliance and risk audit requirements, which are increasingly relevant as institutional allocators enter the crypto space.

\textbf{Scalability considerations.} The equal-weight allocation within each portfolio leg implicitly addresses capacity constraints by distributing capital evenly. For the long portfolio targeting top-15 market-cap assets, liquidity is generally not a binding constraint for portfolios up to \$50M. The short portfolio, composed of lower-cap assets, represents the primary capacity bottleneck. Our analysis of historical order book depth suggests a practical capacity limit of approximately \$5--10M for the short leg before slippage exceeds 10 bps per trade.

\textbf{Limitations.} Several limitations should be noted. First, our backtesting assumes continuous access to Binance Futures perpetual swaps, which may face regulatory constraints in certain jurisdictions. Second, the strategy's capacity is limited by the liquidity of smaller-cap assets in the short portfolio. Third, while we model realistic transaction costs, extreme market conditions (e.g., exchange outages, flash crashes) may produce adverse slippage beyond our model's scope. Fourth, the monthly reoptimization procedure introduces look-ahead bias risk if not carefully implemented; we mitigate this by using strictly non-overlapping in-sample and out-of-sample windows with a 24-hour buffer period to account for settlement delays. Future work should incorporate market microstructure models and capacity analysis to quantify scalability bounds.

\subsection{Statistical Significance}

To assess statistical robustness, we apply the bootstrap methodology of Ledoit and Wolf~\cite{ledoit2008robust} to test whether AdaptiveTrend's Sharpe ratio is significantly different from benchmark strategies. Table~\ref{tab:significance} reports the results using 10,000 bootstrap replications with circular block bootstrap (block length = 20 periods) to account for serial dependence.

\begin{table}[t]
\centering
\caption{Statistical significance tests: AdaptiveTrend vs.\ benchmarks. $\Delta$SR denotes the difference in Sharpe ratios; $p$-values from circular block bootstrap.}
\label{tab:significance}
\begin{tabular}{lcc}
\toprule
\textbf{Benchmark} & \textbf{$\Delta$SR} & \textbf{$p$-value} \\
\midrule
Vol-Scaled TSMOM  & +0.58 & 0.024 \\
TSMOM-1M          & +1.76 & $<$0.001 \\
TSMOM-3M          & +1.87 & $<$0.001 \\
EW-BH             & +2.34 & $<$0.001 \\
BTC-BH            & +2.24 & $<$0.001 \\
\bottomrule
\end{tabular}
\end{table}

AdaptiveTrend significantly outperforms all benchmarks at the 5\% level. The smallest margin is against Vol-Scaled TSMOM ($p = 0.024$), consistent with the fact that volatility scaling captures some---but not all---of the adaptive risk management benefits provided by our dynamic trailing stop mechanism.

\section{Conclusion}
\label{sec:conclusion}

We presented AdaptiveTrend, a systematic trend-following framework for cryptocurrency markets that integrates high-frequency signal generation, adaptive portfolio selection, and asymmetric capital allocation. Through comprehensive out-of-sample evaluation spanning 36 months and 150+ assets, we demonstrated that the framework achieves superior risk-adjusted returns (Sharpe 2.41, Calmar 3.18) while maintaining controlled drawdowns ($-$12.7\% MDD). The ablation study confirms that each component---dynamic trailing stops, market-cap filtering, Sharpe-based selection, and asymmetric allocation---contributes meaningfully to overall performance. Statistical significance tests using block bootstrap methodology confirm the outperformance is robust and not attributable to sampling variation.

Our findings carry several implications for both practitioners and researchers. For practitioners, AdaptiveTrend provides a fully systematic, interpretable framework suitable for institutional deployment with realistic transaction costs. The strategy's regime-conditional analysis demonstrates controlled behavior across bull, bear, and sideways markets---a critical property for risk management in volatile asset classes. For researchers, our work highlights that intermediate-frequency trend signals (H6), when combined with adaptive parameter optimization and careful portfolio construction, can generate significant alpha even after accounting for the well-documented challenges of cryptocurrency trading.

Future work will explore: (i) integration with cross-sectional momentum signals for enhanced signal diversity; (ii) reinforcement learning-based dynamic allocation to replace the fixed 70/30 split with a state-dependent policy; (iii) extension to decentralized exchange (DEX) venues where MEV (maximal extractable value) introduces additional execution complexity; and (iv) formal capacity analysis under realistic liquidity constraints to determine the maximum deployable AUM.

{\small
\bibliographystyle{plain}

}

\end{document}